\documentclass[runningheads]{svmult}
\usepackage{makeidx}   % allows index generation
\usepackage{graphicx}  % standard LaTeX graphics tool
\usepackage{subeqnar}  % subnumbers individual equations
\usepackage{multicol}  % used for the two-column index
\usepackage{cropmark} % cropmarks for pages without
\usepackage{physprbb}  % centered layout of diverse elements, etc.
\begin{document}
\title*{Neutrino Data, CP violation and Cosmological Implications}
\toctitle{Neutrino Data, CP violation and Cosmological Implications}
% allows explicit linebreak for the table of content
%
%
\titlerunning{Leptonic CP Violation}
% allows abbreviation of title, if the full title is too long
% to fit in the running head
%
\author{M. N. Rebelo
%\inst{1}
%\and Roger Temam\inst{2}
%\and Jeffrey Dean\inst{2}
%\and David Grove\inst{1}
%\and Craig Chambers\inst{2}
%\and Kim~B.~Bruce\inst{2}
%\and Elsa Bertino\inst{1}}
%
\authorrunning{M. N. Rebelo}
% if there are more than two authors,
% please abbreviate author list for running head
%
%
\institute{Departamento de F\'{\i}sica and Grupo Te{\'o}rico 
de F{\'\i}sica de Part{\'\i}culas, 
Instituto Superior T\'{e}cnico, Av. Rovisco Pais, 1049-001 Lisboa,
Portugal}}

\maketitle              % typesets the title of the contribution

\begin{abstract}
Recent experimental data provides evidence for neutrino masses
and leads to the possibility of leptonic mixing and CP violation.
\index{abstract} In this work special attention is dedicated
to CP violation in the leptonic sector both at low
and at high energies in the framework of seesaw with only three
righthanded neutrinos added to the Lagrangean of the Standard Model. 
It is shown that leptogenesis is a
possible and likely explanation for the observed baryon asymmetry of the
universe. In this case neutrino masses are constrained yet CP
violating phases at low energies can only be related to
CP violation at high energies in the context of specific models. 
\end{abstract}

\section{Introduction}
Recent evidence for neutrino masses available from
experiments with solar
\cite{Fukuda:2002pe}, \cite{Ahmad:2002ka}
and atmospheric neutrinos \cite{Nakaya:2002ki}, reactor experiments
\cite{Apollonio:1999ae}, \cite{Eguchi:2002dm} 
neutrinoless double beta decay searches \cite{kkg} 
astrophysics and cosmology \cite{Bennett:2003bz} \cite{Spergel:2003cb}
entails the possibility
of leptonic mixing and CP violation both at low energies
and at high energies. CP violation in the leptonic sector 
can have important cosmological implications playing a r\^ ole
in the generation of the observed baryon number asymmetry of
the universe (BAU) through leptogenesis \cite{Fukugita:1986hr}.

There are several possible ways of generating neutrino masses
in the context of minimal 
extensions of the Standard $SU(2) \times U(1)$ Model. 
The most 
straightforward one is the simple extension of the 
Standard Model (SM) by including one
righthanded neutrino per generation. In this case the number
of fermionic degrees of freedom for neutrinos equals those of all 
other fermions in the theory, this fact may be viewed as
adding elegance to the theory. This is the framework
on which the work presented in this talk is based. Special
emphasis will be given to general results obtained 
in  \cite{Branco:2001pq} and
\cite{Rebelo:2002wj}
and at the same time
an attempt is made to present some recent important
results obtained by other authors.
It should be pointed out that it is also possible to generate
neutrino masses in such a framework without requiring
the number of righthanded and lefthanded neutrino fields
to be equal.

It is well known that such an extension of the SM allows for
the seesaw mechanism \cite{see} to operate giving rise
to three light and three heavy neutrinos of Majorana
character as well as leptonic mixing and the possibility
of CP violation in the couplings of these neutrinos
to the charged leptons. The seesaw mechanism also
provides a natural explanation for the smallness of
neutrino masses. In this framework one of the most plausible
scenarios for the generation of BAU is the leptogenesis
mechanism where a CP asymmetry generated through the
out-of-equilibrium L-violating decays of the heavy 
Majorana neutrinos leads to a lepton asymmetry which
is subsequently transformed into a baryon asymmetry 
by (B+L)-violating sphaleron processes \cite{Kuzmin:1985mm}.

The possibility of CP violation in the leptonic sector 
both at low and high energies, i.e., in the charged current
couplings of heavy neutrinos -- with implications for
leptogenesis -- and the charged current couplings of light
neutrinos -- with implications for low energy phenomenology
that may possibly be observed in future experiments --
raises the important question of whether it is possible to
establish a direct connection between these two phenomena.
This question is of
special relevance and is related to another important
one: If indeed BAU is generated through leptogenesis
how can it be proved? In the discussion that follows
it will become clear that this connection cannot 
be established in general it can only be established in special
frameworks. This was shown in  \cite{Branco:2001pq}, 
by making use of a special parametrization which made
explicit the fact that it is possible to have low energy
CP violation without CP violation at high energies
(meaning no leptogenesis). In addition it was shown 
in  \cite{Rebelo:2002wj} that viable leptogenesis is
possible without low energy CP violation, i.e., no CP
violation at low energies resulting either from Dirac
or from Majorana phases. Several authors have addressed
the same question in the context of specific models \cite{varios}.

In the next section we introduce the general framework and 
give the number of independent CP violating phases 
present in the Lagrangean. We show how 
these phases can be parametrized, for three generations,
still in a weak basis and we also indicate how they
appear in the physical basis, through the seesaw
mechanism, where three phases are relevant for
CP violation at low energies. 
In section three we briefly present the conditions
for viable leptogenesis with special emphasis
in the case of hierarchical heavy neutrinos. We show that
for three generations there are three CP violating
phases on which leptogenesis depends.
In section four we comment on the connection between
CP violation at high energies and at low energies.
Section five shows how to build weak basis 
invariant conditions allowing
to determine whether a particular Lagrangean does violate CP
without the need to go to the physical basis. Section six 
contains the conclusions.

\section{General Framework and Seesaw}
After spontaneous symmetry breaking, the leptonic
mass term for the minimal extension of the SM, which consists of 
adding to the standard spectrum one right-handed neutrino 
per generation, can be written as:
\begin{eqnarray}
{\cal L}_m  &=& -\left[ \overline{{\nu}_{L}^0} m \nu_{R}^0 +
\frac{1}{2} \nu_{R}^{0T} C M \nu_{R}^0+
\overline{l_L^0} m_l l_R^0 \right] + 
{\rm h. c.} = \nonumber \\
&=& - \left[ \frac{1}{2}  n_{L}^{T} C {\cal M}^* n_L +
\overline{l_L^0} m_l l_R^0 \right] + {\rm h. c.}
\label{lm}
\end{eqnarray}
where $m$, $M$ and $m_l$ denote the neutrino Dirac mass matrix,
the right-handed neutrino Majorana mass matrix and the charged
lepton mass matrix, respectively, and
$n_L = ({\nu}_{L}^0, {(\nu_R^0)}^c)$ (should be interpreted
as a column matrix). In this minimal
extension of the SM a term of the form
$\frac{1}{2} \nu_{L}^{0T} C m_L \nu_{L}^0$
does not appear in the Lagrangean and the matrix ${\cal M}$
is given by:
\begin{equation}
{\cal M}= \left(\begin{array}{cc}
0 & m \\
m^T & M \end{array}\right) \label{calm}
\end{equation}
with a zero entry on the (11) block. The right-handed
Majorana mass term is $SU(2) \times U(1)$
invariant, consequently it can have a value much above the 
scale $v$ of the electroweak symmetry breaking, thus leading
to the seesaw mechanism.

The number of independent CP violating phases was
identified for this case \cite{Endoh:2000hc} as being
equal to $n(n-1)$ with $n$ the number of generations.
For three generations this number equals six.
In the most general case, without imposing $m_L$
equal to zero and  with $n$ lefthanded neutrino
fields, $n^\prime$ righthanded neutrino fields 
this number is given  \cite{Branco:gr} by $nn^\prime
+ \frac{n(n-1)}{2}$.

CP violation may be analysed either in a weak basis (WB)
or in the physical basis. It is always possible to
choose a WB where the matrices $M$ and $m_l$ are 
simultaneously diagonalized, in this WB all CP violating 
phases appear in the matrix $m$. The matrix $m$ can be written 
without loss of generality as the product of 
a unitary times a Hermitian matrix (polar decomposition)
in this case it is clear how these six independent
CP violating phases may appear in $m$:
\begin{equation}
m=U H = P_{\gamma}^\dagger
{\hat U_{\varrho}} P_{\tau} {P_{\beta}}^\dagger
{\hat H_{\sigma}} P_{\beta} \; .
\label{uhm}
\end{equation}
The matrices $P$ are diagonal unitary, 
in general $P_{\gamma}$ can have
three phases in the diagonal which can be rotated
away through a redefinition of the lefthanded leptons,
$P_{\tau}$ and $P_{\beta}$ only have two phases each, 
$\hat U_{\varrho}$ is a general unitary matrix with
only one phase left, $\hat H_{\sigma}$ is a Hermitian
matrix with two of its phases factored out. After
rotating away  $P_{\gamma}$ we are left with
\begin{equation}
m={\hat U_{\varrho}} P_{\alpha} {\hat H_{\sigma}} P_{\beta} 
\label{upy}
\end{equation}
with six phases $\varrho$, $\alpha_1$, $\alpha_2$, $\sigma$,
$\beta_1$ and $\beta_2$  which cannot be eliminated.

In order to go to the physical basis let us start
from the WB where $m_{l}$ is already diagonal.
The neutrino mass matrix $\cal M$
is diagonalized by the transformation:
\begin{equation}
V^T {\cal M}^* V = \cal D \; , \label{dgm}
\end{equation}
where ${\cal D} ={\rm diag} (m_{\nu_1}, m_{\nu_2}, m_{\nu_3},
M_{\nu_1}, M_{\nu_2}, M_{\nu_3})$,
with $m_{\nu_i}$ and $M_{\nu_i}$ denoting the physical
masses of the light and heavy Majorana neutrinos, respectively. It is
convenient to write $V$ and $\cal D$ in the following block form:
\begin{eqnarray}
V= \left (\begin{array}{cc}
K & R \\
S & T \end{array}\right) ; \ \ \
{\cal D}=\left(\begin{array}{cc}
d & 0 \\
0 & D \end{array}\right) \; .
\end{eqnarray}
From (\ref{dgm}) and assuming the scale of $M$ much
higher than that of $v$,
one obtains, to an excellent approximation:
\begin{equation}
-K^\dagger m \frac{1}{M} m^T K^* =d \; , \label{14}
\end{equation}
together with the following exact relation:
\begin{equation}
R=m T^* D^{-1} \; . \label{exa}
\end{equation}
In the WB where the right-handed Majorana neutrino mass
is also diagonal, it then follows, to an excellent 
approximation, that:
\begin{equation}
R=m D^{-1} \; .  \label{app}
\end{equation}
Equation (\ref{14}) is the usual seesaw formula with $K$
a unitary matrix.
The neutrino weak-eigenstates are related to the mass eigenstates by:
\begin{equation}
{\nu^0_i}_L= V_{i \alpha} {\nu_{\alpha}}_L=(K, R)
\left(\begin{array}{c}
{\nu_i}_L  \\
{N_i}_L \end{array} \right) \quad \left(\begin{array}{c} i=1,2,3 \\
\alpha=1,2,...6 \end{array} \right) \; ,
\label{15}
\end{equation}
and thus the leptonic charged-current interactions are given by:
\begin{equation}
- \frac{g}{\sqrt{2}} \left( \overline{l_{iL}} \gamma_{\mu} K_{ij}
{\nu_j}_L +
\overline{l_{iL}} \gamma_{\mu} R_{ij} {N_j}_L \right) W^{\mu}
+{\rm h.c.}
\label{16}
\end{equation}
From  (\ref{15}), (\ref{16}) we see that $K$ and $R$ give the
charged-current couplings of charged leptons to the light
neutrinos $\nu_j$ and to the heavy
neutrinos $N_j$, respectively. The unitary
matrix $K$, which contains all
the information about CP violation at low energies, 
can be parametrized as:
\begin{equation}
K= P_{\xi} {\hat U_{\delta}} P_{\theta}
\ \ \ \longrightarrow \ \ \ {\hat U_{\delta}} P_{\theta}
\label{kkk}
\end{equation}
with $P_{\xi}={\rm diag}\left(\exp(i\xi_1),\exp(i\xi_2),\exp(i\xi_3)
\right)$, and 
$P_\theta ={\rm diag}(1, \exp(i\theta_1) \exp(i\theta_2))$
leaving ${\hat U_{\delta}}$ with only one phase as in the case of the
Cabibbo, Kobayashi and Maskawa matrix. 
Since $ P_{\xi}$ can still be rotated away
by a redefinition of the charged leptonic fields, $K$ is left
with three CP-violating phases, one of Dirac type $\varrho$ and two
of Majorana character $\theta_1$ and $\theta_2$. The matrix $K$
is the Maki, Nakagawa and Sakata mixing matrix  \cite{Maki:1962mu}.
The matrix $R$ is of relevance for leptogenesis even though the
out of equilibrium decay of the heavy Majorana neutrinos 
responsible for the generation of $L \neq 0$ occurs in the
symmetric phase, since it can be related to the inicial 
Yukawa couplings and the masses of the heavy neutrinos through 
(\ref{exa}). It is clear from 
(\ref{app}) that its entries
are suppressed by $\frac{v}{M}$.

\section{Comment on General Conditions for Leptogenesis}
In this section, we identify the CP violating phases relevant for
leptogenesis, obtained through the out of equilibrium decay of heavy
Majorana neutrinos.
We have seen that there are six independent CP violating 
phases in the Lagrangean.
Next we show which of these six independent CP violating
phases contribute to lepton number asymmetry.
Leptogenesis strongly depends on the masses of the
heavy neutrinos and requires different conditions
depending on whether these masses are hierarchical \cite{tres}
or close to degenerate \cite{Pilaftsis:2003gt}. Thermal
leptogenesis in the case of hierarchical heavy neutrinos
only depends on
four parameters \cite{Buchmuller}: 
the mass $M_{1}$ of the ligthest heavy neutrino
together with the corresponding CP asymmetry $\varepsilon _{N_{1}}$ in their
decays, as well as the effective neutrino mass $\widetilde{m_{1}}$ defined
as 
\begin{equation}
\widetilde{m_{1}}=(m^{\dagger }m)_{11}/M_{1}  \label{mtil}
\end{equation}
in the weak basis where $M$ is diagonal real and positive, and finally, the
sum of all light neutrino mass squared, 
${\overline{m}}^{2}=m_{1}^{2}+m_{2}^{2}+m_{3}^{2}$, 
which controls an important class of
washout processes.

The computation of the lepton-number asymmetry, in this extension 
of the SM, resulting from the decay of a heavy Majorana neutrino $N^j$
into charged leptons $l_i^\pm$ ($i$ = e, $\mu$ , $\tau$) leads to
\cite{sym} :
\begin{eqnarray}
\varepsilon _{N_{j}}
&=& \frac{g^2}{{M_W}^2} \sum_{k \ne j} \left[
{\rm Im} \left((m^\dagger m)_{jk} (m^\dagger m)_{jk} \right)
\frac{1}{16 \pi} \left(I(x_k)+ \frac{\sqrt{x_k}}{1-x_k} \right)
\right]
\frac{1}{(m^\dagger m)_{jj}}   \nonumber \\
&=& \frac{g^2}{{M_W}^2} \sum_{k \ne j} \left[ (M_k)^2
{\rm Im} \left((R^\dagger R)_{jk} (R^\dagger R)_{jk} \right)
\frac{1}{16 \pi} \left(I(x_k)+ \frac{\sqrt{x_k}}{1-x_k} \right)
\right]
\frac{1}{(R^\dagger R)_{jj}} \nonumber \\
\label{rmy}
\end{eqnarray}
with the lepton-number asymmetry from the $j$ heavy Majorana
particle, $\varepsilon _{N_{j}}$, defined in terms of the 
family number asymmetry 
$\Delta {A^j}_i={N^j}_i-{{\overline{N}}^j}_i$ by :
\begin{equation}
\varepsilon _{N_{j}}
= \frac{\sum_i \Delta {A^j}_i}{\sum_i \left({N^j}_i +
\overline{N^j}_i \right)}
\label{jad}
\end{equation}
$M_k$ are the heavy neutrino masses,
the variable $x_k$
is defined as  $x_k=\frac{{M_k}^2}{{M_j}^2}$ and
$ I(x_k)=\sqrt{x_k} \left(1+(1+x_k) \log(\frac{x_k}{1+x_k}) \right)$,
the sum in $i$ runs over the three flavours
$i$ = e $\mu$ $\tau$.
From (\ref{rmy}) it can be seen that the lepton-number
asymmetry is only sensitive to the CP-violating phases
appearing in $m^\dagger m$ in the WB, where $M$ and $m_l$
are diagonal (or equivalently in $R^\dagger R$).
Making use of the parametrization given by (\ref{upy})
it becomes clear that leptogenesis is only sensitive to
the phases $\beta_1$, $\beta_2$ and $\sigma$. If these
phases are zero $m^\dagger m$ is real and no lepton number
asymmetry is generated through the decay of heavy
Majorana neutrinos.

Successful leptogenesis would require 
$\varepsilon _{N_{1}} $ of order $10^{-8}$, if
washout processes could be neglected, 
in order reproduce the observed ratio of baryons
to photons which is given by \cite{Bennett:2003bz}:
\begin{equation}
\frac{n_{B}}{n_{\gamma}}= (6.1 ^{+0.3}_{-0.2}) \times 10^{-10} \; .
\end{equation}
The computation of the effect of washout processes
requires the integration of the full set of
Boltzmann equations.
Leptogenesis is a nonequilibrium process which takes 
place at temperatures $T\sim M_{1}$. This imposes an upper 
bound on the effective neutrino
mass $\widetilde{m_{1}}$  given by the ``equilibrium neutrino
mass'' \cite{Kolb:vq} \cite{Fischler:1990gn} 
\cite{Buchmuller:1992qc}: 
\begin{equation}
m_{*}=\frac{16\pi ^{5/2}}{3\sqrt{5}}g_{*}^{1/2}\frac{v^{2}}{M_{pl}}\simeq
10^{-3}\mbox{eV}  \label{enm}
\end{equation}
where $M_{pl}$ is the Planck mass 
($M_{pl}=1.2\times 10^{19}$ Gev). The sum
of all neutrino mass squared ${\overline{m}}^{2}$ 
is constrained, to be below 0.21 eV 
\cite{anterior}. Which 
implies an upper bound on all light neutrinos masses of 0.12 eV.
Furthermore relaxing this upper bound to 0.4 eV already
requires strong degeneracy of the heavy neutrinos \cite{anterior}.
It is interesting to note that these bounds are
compatible  \cite{Klapdor-Kleingrothaus:2003yg}
with the present constraints on $| <m> |$
defined by:
\begin{equation}
| <m> | \equiv | m_{1} K_{11}^2 +  m_{2} K_{12}^2 +  
m_{3} K_{13}^2 |
\label{mid}
\end{equation}
obtained from  neutrinoless double beta decay, 
for which  the Heidelberg-Moscow Collaboration gives  \cite{kkg}:
\begin{equation}
| <m> | = (0.05 -- 0.84) \mbox{eV} \; .
\label{hmc}
\end{equation}

The leptogenesesis scenario is an interesting explanation
for BAU in agreement with all experimental data available
at present.

\section{On the Connection between CP Violation at Low and 
High Energies}

The prospects of finding CP-violating effects 
at low energies, for instance in 
future neutrino factories, are extremely exciting.
Yet it is important to notice that leptogenesis
remains in principle a viable scenario even if there is 
no CP violation at low energies \cite{Rebelo:2002wj}, conversely
the observation of CP violation at low energies does not
necessarily imply CP violation at high energies 
\cite{Branco:2001pq}.  

In the previous section it was shown that there is no 
leptogenesis for $\beta_1$, $\beta_2$ and $\sigma$ 
equal to zero. However the matrix $m_{eff}= -m\frac{1}{D}m^{T}$
which is diagonalized by the $V_{MNS}$ matrix,
can still be complex in this case, due to the fact
that there are three additional phases $\varrho$, $\alpha_1$
and $\alpha_2$ in the parametrization of $m$ and
these do not cancel out in $m_{eff}$. Another simple way of
reaching the same conclusion is by noting that any matrix
can be diagonalized through a biunitary transformation
and thus writing the matrix $m$ in the form:
\begin{equation}
m = {U_1}^{\dagger} d_D U_2
\label{xyz}
\end{equation}
with the $U_{i}$ unitary matrices and $d_D$ a diagonal real
matrix. If $U_2$ is real $m^{\dagger }m$ is also real and there is
no leptogenesis. Yet $m_{eff}$ which is given by
\begin{equation}
m_{eff} = -m\frac{1}{D}m^{T} = 
-{U_1}^{\dagger} d_D U_2 D^{-1} {U_2}^T d_D {U_1}^*
\label{uuuu}
\end{equation}
can be a complex matrix, even in the limit of $U_2$  real,
requiring $V_{MNS}$ also complex.

On the other hand from (\ref{14}) we can write 
\begin{equation}
\left( m \frac{1} {\sqrt D } \right) \left( m \frac{1} {\sqrt D } \right)^T =
(i K {\sqrt d})(i K {\sqrt d})^T
\label{mul} 
\end{equation}
with ${\sqrt d}$ and ${\sqrt D }$  diagonal real 
matrices such that   ${\sqrt d} {\sqrt d} = d $, 
${\sqrt D } {\sqrt D } = D $ it is clear that it is possible
to choose the matrix $m$ after replacing $K$ 
by the expression given in (\ref{kkk})  as \cite{param}:
\begin{equation}
m= i {\hat U_{\varrho}} P_{\theta} {\sqrt d} O^c {\sqrt D } \; ,
\label{mmm}
\end{equation}
$O^c$ is an orthogonal complex matrix, i.e.  $O^c {O^c}^T = 1 $
but $O^c {O^c}^\dagger \neq 1 $.
Particularizing for $\theta_{1} = \theta_{2} =0$ together with 
$\varrho =0$, there is no CP violation at low energies. Yet 
leptogenesis is sensitive to
the combination $m^\dagger m$,  which is given by:
\begin{equation}
m^\dagger m = {\sqrt D }  {O^c}^\dagger  
d   O^c {\sqrt D } \; ;  
\label{hhh}
\end{equation}
consequently, provided that the combination 
${O^c}^\dagger  d \;   O^c $ is CP-violating, we may have 
leptogenesis even without CP violation at low energies either
of Dirac or Majorana type.  

Equation (\ref{xyz}) is also useful to reach the same conclusion
in a simple way. In this notation 
\begin{equation}
m^\dagger m = {U_2}^{\dagger} {d_D}^2 U_2
\label{lep}
\end{equation}
and viable leptogenesis requires a complex  $U_2$ matrix.
In this case from (\ref{uuuu}) it is clear that, unless one
chooses a special form for the $U_1$ matrix, $m_{eff}$ is
in general complex and there is also CP violation at low energies.
However it is obvious that it is always possible to choose 
$U_1$ such that $V_{MNS}$ is real -- this conclusion follows from
the fact that both $U_1$ and $V_{MNS}$ are  unitary matrices and
appear in adjacent positions in the diagonalization
of $m_{eff}$ so that $U_1$ can be redefined to absorb the 
phases of $V_{MNS}$. 
Furthermore this reasoning shows that, given a model with an   
arbitrary complex matrix $m$, in general one should expect 
manifestations of CP violation both at low and at high energies. 
The connection between these manifestations is model dependent
and has been studied by many authors \cite{varios}. 
It is possible that all CP violating phenomena
in nature have a common origin through a single phase
in the vacuum expectation value of a complex scalar 
field \cite{Branco:2003rt}.

\section{Weak Basis Invariants and CP Violation}
Given a Lagrangean still written in a weak basis it is useful
to be able to analyse whether or not there
is CP violation  without necessarily having to go to the 
physical basis. For that purpose one must write WB invariant
conditions which have to be verified in the case of CP
conservation.

Let us write the
most general CP transformation which leaves the Lagrangean
invariant \cite{Branco:gr}:

\begin{eqnarray}
{\rm CP} l_L ({\rm CP})^{\dagger}&=&U \gamma^0 {\rm C} \overline{l_L}^T
\quad
{\rm CP} l_R({\rm CP})^{\dagger}=V \gamma^0 {\rm C} \overline{l_R}^T
\nonumber \\
{\rm CP} \nu_L ({\rm CP})^{\dagger}&=&U \gamma^0 {\rm C}
\overline{\nu_L}^T \quad
{\rm CP} \nu_R ({\rm CP})^{\dagger}=W \gamma^0 {\rm C} \overline{\nu_R}^T
\label{cp}
\end{eqnarray}
where U, V, W are unitary matrices acting in flavour space
and where for notation simplicity we have dropped here the
superscript 0 in the fermion fields.
Invariance of the mass terms under the above CP transformation,
requires that the following relations have to be satisfied:
\begin{eqnarray}
W^T M W &=&-M^*  \label{cpM} \\
U^{\dagger} m W&=& {m}^*  \label{cpm} \\
U^{\dagger} m_l V&=& {m_l}^* \label{cpml}
\end{eqnarray}
From these equations one obtains:
\begin{equation}
W^{\dagger}h W = h^* \qquad \qquad 
W^{\dagger}H W = H^* \qquad \qquad
U^{\dagger}{h_l} U = {h_l}^* 
\label{wh}
\end{equation}
where $h=m^{\dagger}m$, $H=M^{\dagger}M$ and  $h_l = m_l {m_l}^{\dagger}$.
It can be then readily derived, from (\ref{cpM}),
(\ref{wh}), through multiplications or commutators and applying
traces (and determinants) that CP invariance requires, for instance:
\begin{eqnarray}
I_1 \equiv {\rm Im Tr}[h H M^* h^* M]=0 \nonumber \\
I_2 \equiv {\rm Im Tr}[h H^2 M^* h^* M] = 0  \\
I_3 \equiv {\rm Im Tr}[h H^2 M^* h^* M H] = 0 \nonumber 
\label{i1}
\end{eqnarray}
Since these $I_i$ are WB invariant, they may be evaluated in any convenient WB.
These conditions are sensitive to the phases $\beta_1$, $\beta_1$ and $\sigma$
which are relevant for leptogenesis. Three additional interesting conditions
can be obtained through the substitution of $h$ by ${\bar h}=m^{\dagger} h_l
m$. The strength of CP violation at low energies, observable 
for example through neutrino
oscillations, can be obtained from the following low-energy WB invariant:
\begin{equation}
Tr[h_{ef}, h_l]^3=6i \Delta_{21} \Delta_{32} \Delta_{31}
{\rm Im} \{ (h_{ef})_{12}(h_{ef})_{23}(h_{ef})_{31} \} \label{trc}
\end{equation}
where $h_{ef}=m_{eff}{m_{eff}}^{\dagger}$ and
$\Delta_{21}=({m_{\mu}}^2-{m_e}^2)$ with analogous expressions for
$\Delta_{31}$, $\Delta_{32}$. This invariant is analogous to
the one written for the quark sector in the context of
the standard model in  \cite{Bernabeu:fc} where this technique 
was first applied.
Several different WB invariant conditions, useful 
in the leptonic sector, and for specific models
have been built using the same technique \cite{Branco:gr} \cite{useful}

\section{Conclusions}
Neutrino physics is a very lively subject both theoretically
and experimentally. Several neutrino experiments are under study 
for the near future. From the theoretical point of view
a lot of work is being done on fundamental questions such
as the origin of leptonic masses and mixing \cite{Ma:2003im}.
Neutrino properties have important cosmological implications,
for example, the possibility of leptogenesis. If leptogenesis
is the origin of the observed baryon asymmetry of the
universe this implies constraints on neutrino masses both of light
and of heavy neutrinos.
Leptogenesis is one of the most promissing scenarios,
in part due to the fact that several other alternative
proposals are on the verge of being ruled out.  However  
it is likely that this will 
remain an open question still for some time.

\section*{Acknowledgments}

The author thanks the organizers of 
Beyond 2003 for the warm hospitality at Ringberg Castle
and the stimulating Conference. This
work was partially supported by Funda\c{c}\~{a}o para a Ci\^{e}ncia e a
Tecnologia (FCT, Portugal) through the projects,
POCTI/36288/FIS/2000 and  CFIF--Plurianual (2/91) which 
is partially funded through POCTI (FEDER).

%INDEX%%%%%%%%%%%%%%%%%%%%%%%%%%%%%%%%%%%%%%%%%%%%%%%%%%%%%%%%%%%%%%%
% Please check with the editor of your book whether he plans to
% include a "mutual" subject index - if so, please code your entries
% in the standard syntax. For your own purposes you may print your
% "personal" index by using the following commands:
%
%\clearpage
%\addcontentsline{toc}{section}{Index}
%\flushbottom
%\printindex
%%%%%%%%%%%%%%%%%%%%%%%%%%%%%%%%%%%%%%%%%%%%%%%%%%%%%%%%%%%%%%%%%%%%%

\end{document}